\documentclass[12pt,preprint]{aastex}
\def\hmsh{$^{\textrm{h}}$}
\def\hmsm{$^{\textrm{m}}$}

\def\2phn{\phn\phn}
\def\3phn{\phn\phn\phn}
\def\4phn{\phn\phn\phn\phn}
\def\12phn{\4phn\4phn\4phn}

\begin{document}

\title{Difference Imaging of Lensed Quasar Candidates in the SDSS Supernova Survey Region}
\author{Brian C. Lacki\altaffilmark{1}, Christopher S. Kochanek\altaffilmark{1}, Krzysztof Z. Stanek\altaffilmark{1}, Naohisa Inada\altaffilmark{2}, Masamune Oguri\altaffilmark{3}}
\altaffiltext{1}{Department of Astronomy and the Center for Cosmology and AstroParticle Physics, The Ohio State University}
\altaffiltext{2}{Cosmic Radiation Laboratory, RIKEN (The Physical and Chemical Research Organization), 2-1 Hirosawa, Wako, Saitama 351-0198, Japan}
\altaffiltext{3}{Kavli Institute for Particle Astrophysics and Cosmology, Stanford University, 2575 Sand Hill Road, Menlo Park, CA 94025, USA}
\email{lacki@astronomy.ohio-state.edu}

\begin{abstract}
Difference imaging provides a new way to discover gravitationally lensed quasars because few non-lensed sources will show spatially extended, time variable flux.  We test the method on lens candidates in the Sloan Digital Sky Survey (SDSS) Supernova Survey region from the SDSS Quasar Lens Search (SQLS) and their surrounding fields.  Starting from 20768 sources, including 49 SDSS quasars and 36 candidate lenses/lensed images, we find that 21 sources including 15 SDSS QSOs and 7 candidate lenses/lensed images are non-periodic variable sources.  We can measure the spatial structure of the variable flux for 18 of these sources and identify only one as a non-point source.  This source does not display the compelling spatial structure of the variable flux of known lensed quasars, so we reject it as a lens candidate.  None of the lens candidates from the SQLS survive our cuts.  Given our effective survey area of order 0.71 square degrees, this indicates a false positive rate of order one per square degree for the method.  The fraction of quasars not found to be variable and the false positive rate should both fall if we analyze the full, later data releases for the SDSS fields.  While application of the method to the SDSS is limited by the resolution, depth, and sampling of the survey, several future surveys such as Pan-STARRS, LSST, and SNAP will avoid these limitations.
\end{abstract}


\section{Introduction}
Gravitational lensing has many applications, from exoplanet searches to large scale structure (see the reviews by Kochanek, Schneider, and Wambsganss in the Saas Fe lectures, 2006).  Galaxy scale lenses can be used to study the dark matter mass profile of the lens galaxy \citep[e.g.][]{Kochanek91,Rusin05,Koopmans06,Jiang07}.  When the background source is a quasar, microlensing by individual stars in the lens galaxy can be used to probe the structure of the quasar's accretion disk \citep{Poindexter07,Morgan07} and broad line regions \citep{Eigenbrod07}.  Galaxy scale lenses can also constrain cosmological parameters \citep{Refsdal64,Oguri07}.  Unfortunately, most applications require large samples of gravitational lenses to be competitive with other methods, while fewer than a hundred lensed quasars are known.  Moreover, these lenses were discovered using different methods with different biases, which is especially problematic if homogeneous samples with well-understood selection functions are needed \citep[see][]{Kochanek04}.  

The known lenses were found by their morphological structure or the presence of higher redshift features in the spectrum of a lower redshift galaxy.  Morphological surveys examine optical or radio images of quasars for evidence that they are lensed.  This works best for point-like sources like optical quasars, and flat spectrum radio sources.  The two largest searches for lensed AGN are the Cosmic Lens All-Sky Survey of flat spectrum radio sources \citep{Myers03,Browne03} and the SDSS Quasar Lens Search \citep[SQLS;][]{Oguri06}.  The radio samples are limited by the number of sufficiently bright radio sources and the difficulties in obtaining source redshifts.  The optical quasar samples suffer from confusion from stars and galaxies and the effects of color changes, either from the starlight or dust in the lenses, on sample selection.  The spectroscopic method made several serendipitous discoveries of lensed quasars, such as Q2237+0305 \citep{Huchra85} and SDSSJ090334.92+502819.2 \citep{Johnston03}.  Following theoretical investigations \citep[e.g.][]{Kochanek92,Miralda92,Mortlock00,Mortlock01}, the SLACS survey \citep{Bolton04,Bolton06} used the spectroscopic method on massive early-type galaxies in the SDSS to identify $\sim$50 candidate lensed starforming galaxies, and confirmed 19 of 29 using Hubble Space Telescope images.  The spectroscopic method is limited by its low yield (about 1 lens candidate per 1000 luminous red galaxy spectra) and biases in its mass range (due to the size of the spectroscopic aperture).  Moreover, by selecting targets based on the properties of the lens galaxy, these lenses are mainly useful for studying the lens galaxies rather than for cosmology \citep[see][]{Kochanek04}.

\citet{Kochanek06} proposed a new method to find lensed quasars based on difference imaging.  Difference imaging, also known as image subtraction, is a way to measure the variable intensity in a region of sky \citep{Tomaney96,Alard98,Alard00}.  It has been used to search for a broad range of variable sources, such as planets transiting stars \citep[e.g.][]{Hartman04}, microlenses \citep[e.g.][]{Alcock99,Wozniak00}, and supernovae \citep[e.g. the Sloan Digital Sky Survey II Supernova Survey,][]{Sako07}.  In difference imaging, a reference image is made by averaging a set of the best images for a field.  Then, for each epoch of observation, a difference image is created by subtracting the reference image convolved to the PSF and flux scale of that epoch.  If a source is variable, there will be a flux residual in the difference image corresponding to the variability of the source.

In difference images, lensed quasars are recognizable because they consist of multiple variable images that are close together.  Since most quasars are variable, with 60\% having $\sigma_g \ge 0.05$ variability over two years \citep[e.g.][]{Sesar07}, each of these images is variable.  The level of variability is then enhanced by microlensing of the quasar images by the stars in the lens galaxy \citep[see the review by][]{Wambsganss06}.  Thus, lensed quasars look like compact clusters of variable sources or extended variable sources, which allows us to easily search for lensed quasars in difference images.  The number of false positives from pairs of variable stars, pairs of quasars (related or not), variable star-quasar blends, and supernovae-AGN pairs is expected to be low \citep{Kochanek06}.  Candidates found this way can be confirmed by light curve analysis, since each image will have the same intrinsic variability, but with time delays between the images and some additional uncorrelated variability from microlensing by the stars of the lens galaxy \citep[e.g.][]{Pindor05}.  Difference imaging can also resolve variable lensed sources blended with non-variable sources, such as the lens galaxies.  This is important because, as we search for fainter lensed quasars, contamination from the lens galaxy becomes a steadily greater problem.  Tests of image subtraction on the lens Q2237+0305 show that the method can find quasars even when they are buried by the flux of an extraordinarily bright foreground galaxy \citep{Kochanek06}.

The forthcoming, large scale synoptic surveys, like LSST \citep{Tyson02} and Pan-STARRS \citep{Kaiser04}, will be ideal for this method, since they will cover large areas of the sky, sample variable sources frequently, and have deep magnitude limits.  In the meantime, the SDSS II Supernova Survey \citep{Frieman07,Sako07} is the best survey currently available for our goals.  Intended to find supernovae for dark energy studies, the Supernova Survey repeatedly images SDSS Stripe 82, a 2.5\degr~wide swath along the celestial equator, covering 300 square degrees and stretching across the Southern Galactic Cap from right ascension 300\degr~to right ascension 60\degr~\citep{Frieman07}.  Ten to twenty public epochs were available for a given field when we started this project, with the number rising to about forty in the most recent releases \citep{Adelman-McCarthy07}.  

We applied the method outlined in \citet{Kochanek06} to the SDSS Supernova Survey fields that contained 26 previously identified candidates for gravitational lenses, with 39 total components, from the initial SQLS candidate list.  These include 15 candidates with 24 components in the main statistical sample \citep{Inada07}, and 11 additional candidate systems, with 15 images, that were outside the final selection criteria.  These candidates are listed in Table~\ref{table:InadaLenses}.  We also checked to see if any pair of lens images in a candidate lens appeared on the difference images, since the images can be well-separated.  The lens candidates had already been rejected for other reasons, such as different spectral energy distributions for the postulated lens images, or the lack of a lens galaxy.  Unfortunately, there are no confirmed lensed quasars in the Supernova Survey region.  We had two goals.  First, to use the variability method as an independent check of the SDSS candidates.  Second, to get a sense of the false positive rate from any other variable sources in the fields.  We outline the method, our approach to source selection, and the results when applied to targets in \S 2.  We discuss the variable sources in the lens candidate fields in \S 3, including quasars (\S 3.2) and an extended variable source (\S 3.3).  Finally, we conclude in \S 4 by discussing prospects for finding gravitational lenses in future surveys with difference imaging.

\section{Procedure}
\subsection{Preparation and Image Subtraction}
For each field around the lens candidates, we downloaded the r-band images from the SDSS Data Server, which had epochs through 2005.  We chose r-band because the r-band SDSS images have great effective depth \citep{York00}.  We then created images centered on the candidate's RA by combining the image containing the candidate with either the previous or next image in the drift scan with an appropriate RA offset.  The RA centered images had the same dimensions and pixel scale as the original SDSS images -- 2048 pixels by 1489 pixels with a pixel scale of 0\farcs396.  The median FWHM seeing in the images was 3.23 pixels (1\farcs3).  We masked data within 10 pixels of bad pixels, because these prevent clean image subtraction.  The bad pixels were usually caused by saturation on the images, so the masked areas differed little from one epoch to the next.  Since the blank areas could not be used for either the reference image or image subtraction, we merged the masks from all observations of a particular field into a common mask.  We also enlarged the masks on the difference images themselves before creating the ``absolute value'' images (see \S 2.2), in order to minimize edge effects.  We lost 15.8\%~of the area covered by the images to masking.

After using ISIS \citep{Alard00} to register the images, we created a reference image.  We used the nine epochs with the best seeing, as measured by PSF size.  If fewer than 18 epochs were available, we instead used only images with better than median seeing.  The median seeing of the reference images was 2.35 pixels, or 0\farcs93.  Finally, we created the difference images.  We used ISIS to align the images, determine the convolution kernel needed to convert the PSF of the reference image to that for each epoch, and then carry out the image subtraction.  In some epochs, the difference image did not seem clean, perhaps because of poor seeing or thin clouds.  To compensate, we eliminated the two epochs for each source with fluxes that were furthest from that source's mean flux, and we did not use images in which the measured magnitude of the sources in the field was on average more than 1 $\sigma$ from the mean.  We limited our structural analysis to sources with peak surface brightnesses $\mu$ in this image with $\mu < \mu_{back} + 0.5$ relative to the local background $\mu_{back}$, where $\mu$ and $\mu_{back}$ are in magnitudes per pixel area.

\subsection{Photometry}
Once we had the difference images, the next step was to select candidate variable sources in each field.  We used ISIS to extract the uncalibrated light curves of sources in the field of each candidate.  While we were not concerned with the absolute flux of the sources, only their variability and their relative brightnesses, we did set an approximate calibration using SDSS stars (point sources) within 3\arcmin~of each quasar.  We used the average offset between the ISIS light curves and the SDSS magnitudes to calibrate each field.  We selected our lens candidates in three steps.

\begin{figure}
\plotone{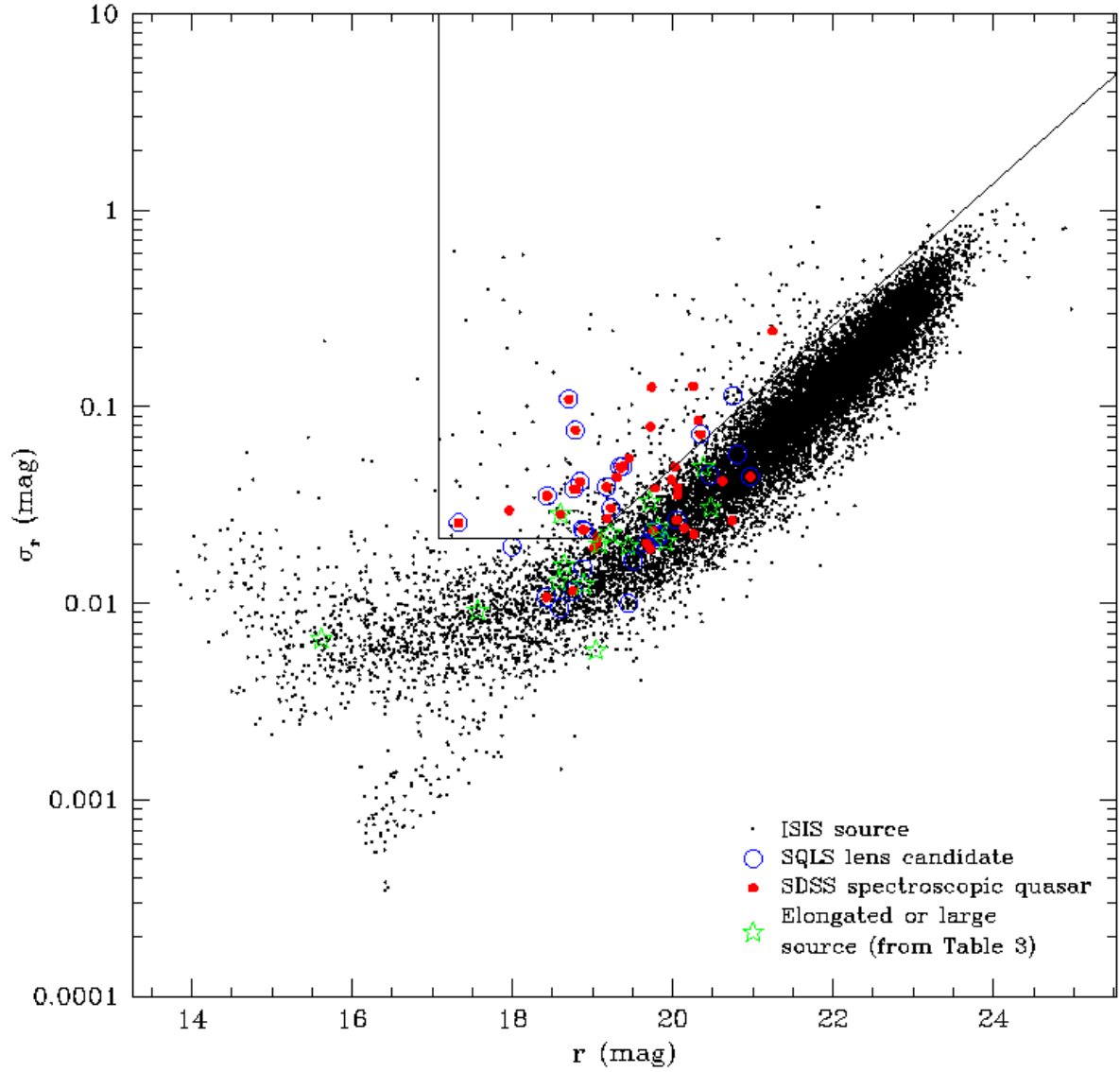}
\caption{The standard deviation of the light curves versus average magnitudes.  Our variability cut is marked by the black line segments -- sources above the boundary are considered to be variable.}
\label{fig:mmVar}
\end{figure}

First, we made a variability cut on the $r$$-$$\sigma_r$ (magnitude$-$magnitude variance) plane.  We excluded sources with more than 30000 counts ($r < 17.1$), because they could be saturated.  We also excluded sources with less than 2\% variability ($\sigma_r < 0.022$) to account for the survey's photometric accuracy.  Finally, we fit a line to the $r$$-$$\textrm{log}~\sigma_r$ relation.  Those sources that had a $\sigma_r$ greater than the 4 $\sigma$ upper bound on the line fit were considered variable.  Quantitatively, we considered sources to be variable only if they satisfied $\textrm{log}_{10}~\sigma_r > 0.362r - 8.56$.   These cuts are shown in Figure~\ref{fig:mmVar}.

Second, quasars have long term, non-periodic variability (LPV) rather than periodic variability, so we identified quasars as objects whose light curves are significantly better fit  as parabolas than constants.  Following \citet{Hartman04}, we used the F-test to gauge the significance of the improvement of a parabolic fit over a constant light curve.  The F-test is based on the ratio
\begin{equation}
F = {\chi^2_p / (N - 4) \over \chi^2_c / (N - 2)}
\end{equation}
where $\chi^2_p / (N - 4)$ is the $\chi^2$ per degree of freedom for the parabolic fit, $\chi^2_c / (N - 2)$ is the $\chi^2$ per degree of freedom of a constant light curve, and $N$ is the number of epochs being fit.  We calculated $F$ after dropping the two epochs most poorly fit by a constant lightcurve, and those epochs with bad images, as described in \S 2.1.  The expected mean of $F$, if $\chi^2_p$ and $\chi^2_c$ are independent chi-square distributions and $\chi^2_c$ has $(N - 2)$ degrees of freedom, is
\begin{equation}
\mu_F = {N - 2 \over N - 4}
\end{equation}
and the expected dispersion in $F$ is
\begin{equation}
\sigma_F^2 = {2 (N - 2)^2 (2N - 8) \over (N - 4)^3 (N - 6)}
\end{equation}
so we use
\begin{equation}
S_F = {F - \mu_F \over \sigma_F}
\end{equation}
to estimate the significance of the improvement from using a parabola.  Since $\chi^2_p$ and $\chi^2_c$ are not independent, with $\chi^2_p \le \chi^2_c$ for each light curve, $\mu_F$ is not the actual mean of the calculated $F$.  It can be shown that $S_F \le 0$ for all sources.  A cut of $S_F \le -0.7$ seemed to work fairly well when we examined the light curves by eye, although the choice was somewhat arbitrary.  Raising the threshold would increase the completeness at the price of introducing more false positives, while lowering it has the opposite effect.  

Finally, to evaluate the structure of the variable flux, we summed the absolute values of the difference images from each epoch.  Variable sources appear as positive peaks in these ``absdiff'' images, whatever their light curves.  We then used SExtractor to identify and characterize the variable sources in these images, focusing on sources that passed our variability and F-test cuts.  We measured the effective half-light radius $R_e$ containing 50\% of the flux.  We also measured the elongation, defined by the ratio of the major axis to the minor axis fit to the source.  Sources that had large effective radii ($R_e > 2.5~\textrm{pixels} = 0\farcs99$) or were elongated ($\epsilon > 1.75$) were considered extended.

\section{Variables in the Fields}
Table~\ref{table:SourceNumbers} lists the number of sources that survive each cut and some alternate combinations of the cuts.  When we consider all sources, after excluding bad images (usually due to low transparency) and dropping two epochs from each light curve, we could only analyze 23 of the 26 fields -- the light curves for the SDSSJ2134$-$0054, SDSSJ2139$-$0114, and SDSSJ2200$-$0107 fields had too few epochs for analysis.  In these 23 fields, which cover an unmasked area of 0.71 square degrees, we can measure light curve statistics, such as $\sigma_r$ and $S_F$, for 15064 objects.  We have statistics for 29 candidate images from 21 candidate lens systems of the SQLS, as well as 44 spectroscopically identified quasars in \citet{Schneider07}.  Seven candidate lens images and 14 spectroscopic quasars could be analyzed in the ``absdiff'' images, with only one spectroscopic quasar being extended.

Figure~\ref{fig:mmVar} shows the mean magnitude and the standard deviation of each light curve, excluding any dropped epochs.  Our sources span the range $14 \la r \la 24$.   The variance shows the usual dependence on magnitude, scaling with the photon noise until saturation effects begin to dominate at $r \la 15$.  At the typical magnitude of an SDSS quasar, $r \sim 19$, the typical variance is of order 0.01 magnitudes.  Variable sources lie above the $r$$-$$\sigma_r$ curve, and we see that these include some, but not all, of the SQLS lens candidates and the other quasars in the field.  If we apply the cuts described in \S2.1, we find 370 variable sources of all kinds.

\begin{figure}
\plotone{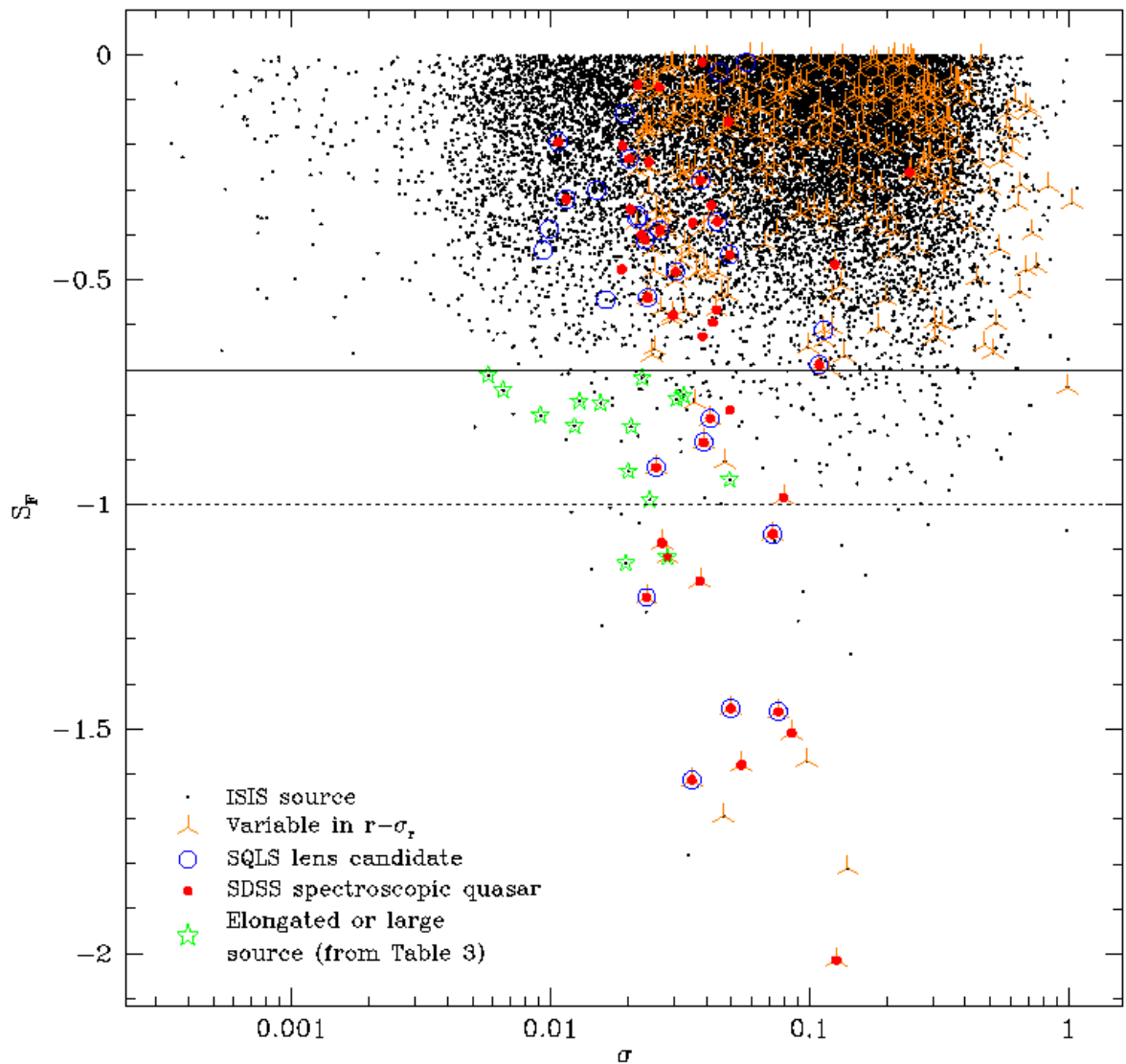}
\caption{The F-test significance $S_F$, as described in \S 2.3, plotted against the standard deviation of the magnitudes.  Our cut of $S_F < -0.7$, indicating long term (LPV-like) behavior, is drawn on the plot as a solid line.  Most of the extended candidates only pass it marginally, with $S_F > -1.0$ (dashed line).}
\label{fig:mVarSF}
\end{figure}

In Figure~\ref{fig:mVarSF}, we plot the distribution of the sources in the space of $S_F$ and $\sigma_r$.  We chose $S_F \le 0.7$ as our cut, since it works well in combination with the cut on the overall variability.  After making those two cuts, we were left with 21 sources, including eight of the SQLS lens candidates and 15 of the SDSS quasars.  A quick inspection of the light curves shows that the F-test does in fact pick out those light curves with a long term trend.  In Figure~\ref{fig:LCs}, we show four light curves for sources that pass the $r$$-$$\sigma_r$ cut, three of which also pass the F-test.  The light curve at upper left (SDSSJ204012.68-003040.36) is one of the sources that did not pass the F-test.  This source is neither a \citet{Schneider07} spectroscopic quasar nor an SQLS lens candidates \citep{Inada07}.  Note the outlier points, of the kind that compelled us to drop some epochs as described in \S2.2.  While the light curve has some variability, especially in the earlier epochs, there is no overall trend, and so it fails the F-test.  The other three light curves pass the F-test and show trends in their variability.  These three variable sources consist of a non-quasar, an SQLS lens candidate \citep{Inada07}, and a spectroscopic quasar which also seems to be extended.  These light curves are for sources that either clearly fail the F-test ($S_F \gg -1$) or clearly pass it ($S_F < -1$) -- for sources near the cut ($S_F \sim -0.7$), the light curves and the F-test results are more ambiguous.  Our F-test cut is generous, in that most of the sources passing it do not clearly pass ($S_F < -1$), as can be seen in Figure~\ref{fig:mVarSF}.

A few of the sources that do pass the variability cuts and the F-test cannot be analyzed on the ``absdiff'' images.  Eighteen of the 21 sources that passed the variability cut and the F-test were measured by SExtractor on the sums of the absolute values of the difference images.  All pass our surface brightness cut of $\mu_{max} < \mu_{back} + 0.5$, where $\mu_{max}$ is the peak surface brightness per pixel after background subtraction, and $\mu_{back}$ is the surface brightness of the background per pixel.  We chose this surface brightness as a measure of whether the SExtractor detection was real, or a spurious detection of noise.  So, we find that 18 sources have significant long period variability and measurable spatial structure to their variability, out of the 15064 initial sources with light curves.

As we survey in Table~\ref{table:SourceNumbers}, we can compare these results with those obtained by omitting either the $r$$-$$\sigma_r$ cut or the F-test.  In the former case, we are left with 208 sources altogether that pass the F-test.  Of theses, 85 were detected by SExtractor on the absdiff images, with only 54 passing our surface brightness cut on the ``absdiff'' images so that they could be analyzed for spatial structure.  This discrepancy, where some SExtractor detections seemed to be spurious, motivated our surface brightness cut.  If, on the other hand, we include the $r$$-$$\sigma_r$ cut but ignore the F-test, then out of the 370 variable sources, 311 were detected on the ``absdiff'' images by SExtractor.  Of these, 158 passed our surface brightness cut and can be analyzed for their spatial structure.  The combination of the two cuts is very effective at reducing the number of candidates relative to the number of quasars known to be present in the fields.

\subsection{The SQLS Lens Candidates}
We looked for both extended SQLS sources and for pairs of sources for the SQLS candidates in the difference images.  Of the original 26 lens candidates in the SDSS Supernova Survey region, we find that some have variable components, but none are extended.  As Table~\ref{table:SourceNumbers} shows, only eight passed the F-test, and seven could be analyzed on the ``absdiff'' images.  None satisfied the criteria given in \S 2.2.  A small separation lensed quasar will appear as a cluster of variable sources on a difference image.  Thirteen of the SQLS candidates are well-separated image pairs, but in all these candidate lenses, at most one image survives the F-test, and the component that passes the tests is always the spectroscopically identified quasar from \citet{Schneider05}, and not the candidate second component selected by color criteria.  Since none of the SQLS lens candidates show extended variability, and since no pair of candidate images have two variable components, none of them are selected by our method either.  We conclude, on independent grounds, that \citet{Inada07} were correct to reject them as lens candidates.

\subsection{Quasars in the Candidate Fields}
Using extended variability to find gravitationally lensed quasars assumes that each quasar is in fact variable enough to appear on difference images.  To address this question, we also examined the images of the 60 SDSS spectroscopic quasars \citep{Schneider07} in the fields of the candidate lenses, including the SQLS lens candidates themselves.  As summarized in Table~\ref{table:SourceNumbers}, 49 of the 60 were detected by ISIS; six of the non-detections were in masked regions, and one, a SQLS lens candidate, was blended with an object identified by \citet{Inada07} as a galaxy.  We were able to perform the least squares fitting necessary for the F-test on 44 quasars.  Of these 44, 15 passed our usual variability cuts and had $S_F \le -0.7$, and we could measure the spatial structure for 14 of these.  So, about one third (32\%) of the quasars for which we have light curve statistics show variability on the difference images.  Only one of the quasars in the candidate lens fields, SDSSJ213245.24+000146.4, was extended using the criteria given in \S 2.  It was not one of the lens candidates in the SQLS, and we discuss it in \S 3.3.

\citet{Sesar07} examined the Supernova Survey region and found that the majority of unresolved quasars listed in an earlier release of the SDSS Quasar Catalog \citep{Schneider05} with $g < 20.5$ were variable.  Specifically, over the two years of the survey, over 60\% of the quasars had variabilities over $\sigma_g > 0.05$, and over 90\% of the quasars had variabilities over $\sigma_g > 0.03$.  When we consider $\sigma_r$ alone, we find less variability than \citet{Sesar07}. Of the forty quasars with statistics and $r < 20.5$, we find only eight (20\%) have variabilities greater than $\sigma_r > 0.05$, and twenty-three (58\%) have variabilities greater than $\sigma_r > 0.03$.  Indeed, Figure~\ref{fig:mmVar} shows that many quasars, especially the fainter ones, are not highly variable.  According to our variability cut on $r$ and $\sigma_r$, twenty-five quasars (57\%) are variable, closer to the proportions in \citet{Sesar07}, but still significantly below 90\%.  The discrepancy is partly caused by our use of r-band images, since the variability is greater in bluer bands \citep{Cutri85,VandenBerk04,Wilhite05}, and partly by the shorter time coverage of our data.

\subsection{Candidate Extended Objects}
\begin{figure}
\plotone{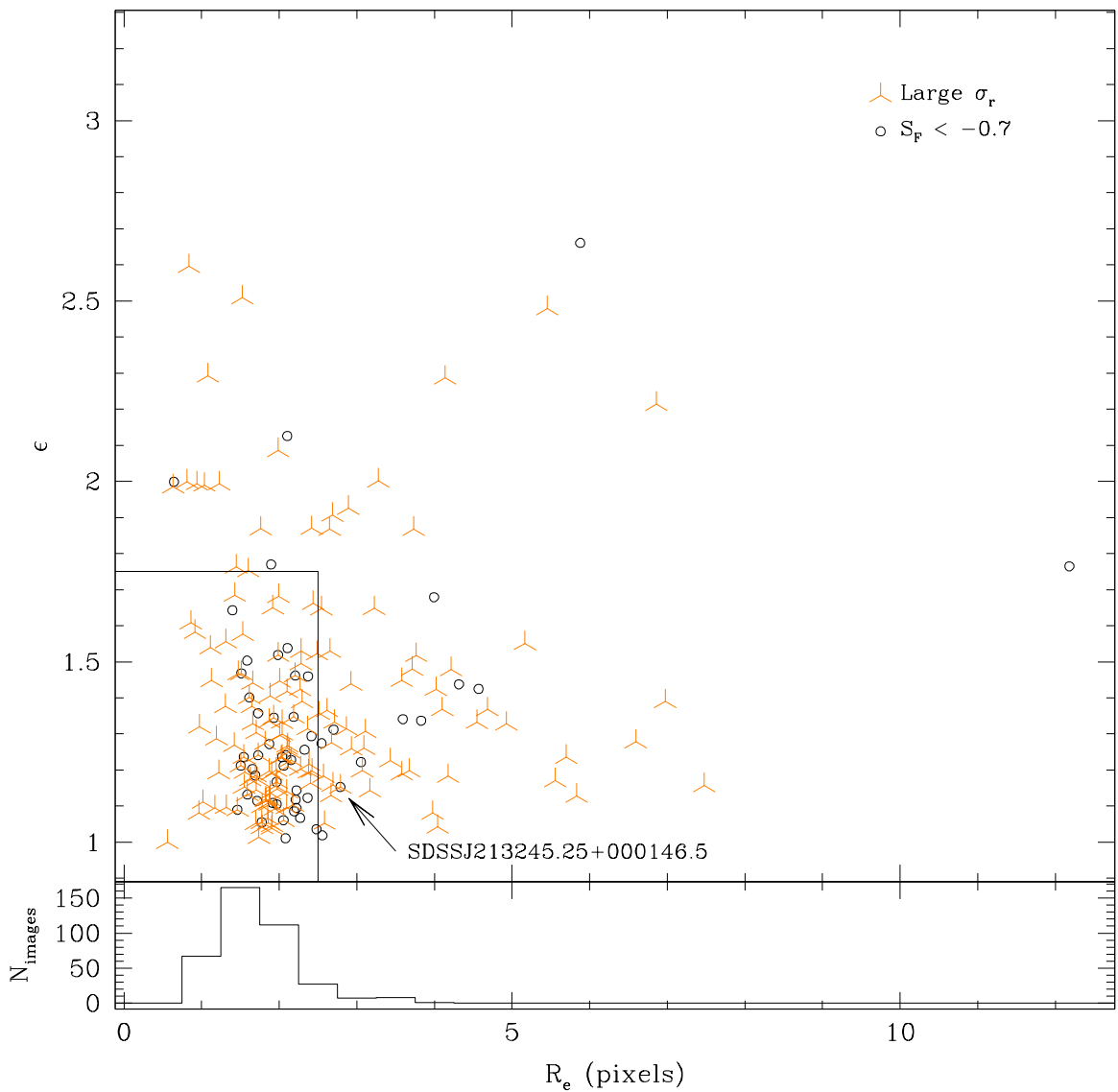}
\caption{Elongation $\epsilon$ and effective radius $R_e$ for sources with peak surface brightness $\mu < \mu_{back} + 0.5$ on the absolute value images, which we considered to be real detections by SExtractor.  Our cut of $\epsilon > 1.75$ or $R_e > 2.5$ pixels is also plotted; those sources above or to the right of the cut were considered extended.  On the bottom, we show a histogram of the seeing ($\textrm{FWHM}/2$) of the original images.}
\label{fig:ReElong}
\end{figure}

Figure~\ref{fig:ReElong} shows the elongations $\epsilon$ (ratios of major to minor axes) and effective (half-light) radii ($R_e$) of the sources that are variable, or have $S_F < -0.7$.  Most sources cluster around $R_e \sim 2~\textrm{pixels}$; since the pixel scale for SDSS is 0\farcs396/pixel, this is consistent with a seeing of about 0\farcs8.  Most also have elongations of $\epsilon \la 1.5$.  There are some outliers, that are either elongated ($\epsilon > 1.75$) or big ($R_e > 2.5$).  For comparison, we measured the seeing of each epoch and compare the $R_e$ distribution to the $\textrm{HWHM} = \textrm{FWHM} / 2$ distribution of the data.  For a Gaussian PSF, the HWHM encloses 50\% of the light.  As can be seen in Figure~\ref{fig:ReElong}, most of the images have measured seeing less than our $R_e$ cut, with a median seeing of $1.6~\textrm{pixels}$~(0\farcs6) and 6.5\% having $\textrm{HWHM} > 2.5~\textrm{pixels}$.  Only one source passed all the cuts, SDSSJ213245.25+000146.5.  

\begin{figure}
\plotone{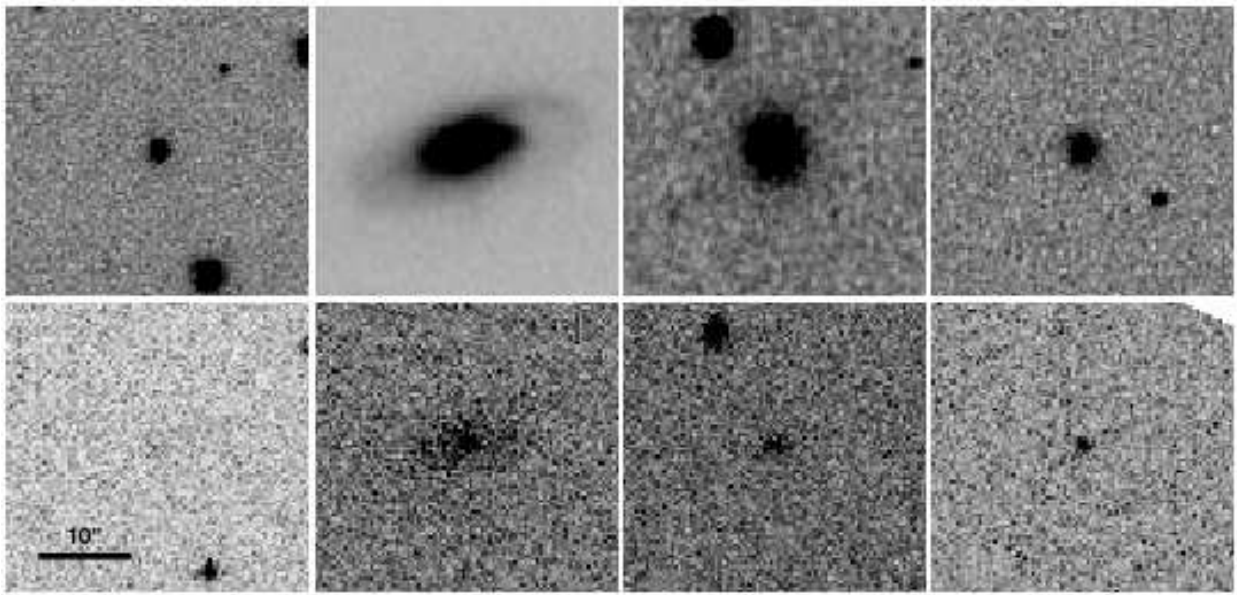}
\caption{Various sources on the reference images (top) and the ``absdiff'' images (bottom).   The images are centered on the objects, and all have the same scale.  At far left is SDSSJ205201.63+000051.6, a quasar with low variability ($\sigma_m = 0.019$ and $S_F = -0.20$), which is not detected by SExtractor in the ``absdiff'' image.  The two middle sources (SDSSJ021247.22+003433.5 at mid-left, SDSSJ021328.83+003030.8 at mid-right) pass the F-test and are extended, but do not pass the $r$$-$$\sigma_r$ cut.  At far-right is SDSSJ213245.25+000146.5, our one candidate extended variable source.}
\label{fig:SourceImages}
\end{figure}

Unlike most variable sources, which only marginally pass the F-test, SDSSJ213245.25+000146.5 clearly passes it with $S_F = -1.117$.  The difference images themselves (Figure~\ref{fig:SourceImages} bottom far-right), however, show no clear structure in the variability.  Furthermore, this source only marginally passed our $R_e$ cut, with $R_e = 2.8~\textrm{pixels}$, and it is not elongated.  This suggests that it is unlikely to actually be a lensed quasar.  SDSSJ213245.25+000146.5 is an X-ray source \citep{Voges00} and one of the quasars in the SDSS Quasar Catalog \citep{Schneider07}.  Since quasars are in fact variable, it is not surprising that it should pass our $r$$-$$\sigma_r$ cut.   However, the quasar is at a redshift of only $z = 0.2342$, while typical lensed quasars are expected to be more distant, with essentially none at such low redshifts \citep[e.g.][]{Mitchell05}.  In addition to showing the quasar, the reference image (as in Figure~\ref{fig:SourceImages}) seems to show the host galaxy of SDSSJ213245.25+000146.5.  \citet{VandenBerk06} considered the source to be resolved and were able to fit galactic components to the spectrum, finding that 2\% of the flux was from the apparent host galaxy.  If SDSSJ213245.25+000146.5 is a lensed quasar, one might expect there to be two sets of spectral lines at different redshifts as in the spectroscopic method of finding lenses \citep[e.g.][]{Bolton06}.  

\section{Conclusion and Future Prospects}
We searched for sources with the spatially extended variability characteristic of gravitationally lensed quasars by applying difference imaging to the Sloan Supernova Survey fields of the SQLS lens candidates and their surroundings.  We found one source, SDSSJ213245.25+000146.5, that passed basic criteria for variability, non-periodic variability, and which appeared extended on its ``absdiff'' image.  However, it did not show the compelling structure seen in difference images of known lensed quasars \citep{Kochanek06} -- it only barely passed the effective radius criterion, suggesting it is not likely to be a true lensed quasar.  Furthermore, it was not among the SQLS lens candidates in the SDSS Supernova Survey region.  Our criteria successfully rejected the SQLS candidates, as had \citet{Inada07} using other criteria.  If used with other lens search techniques, like color or morphological criteria \citep{Oguri06}, difference imaging could substantially reduce the number of false positives, although it would only work for quasars that are variable during the observation period.  Only a third of the quasars detected by ISIS passed our quasar variability tests and could be analyzed for spatial structure in the difference images.  More generally, as a variability survey for lensed quasars covering 0.71 square degrees, we successfully identified one third of the SDSS quasars in the field and had no false positive detections of lenses.

\begin{figure}
\plotone{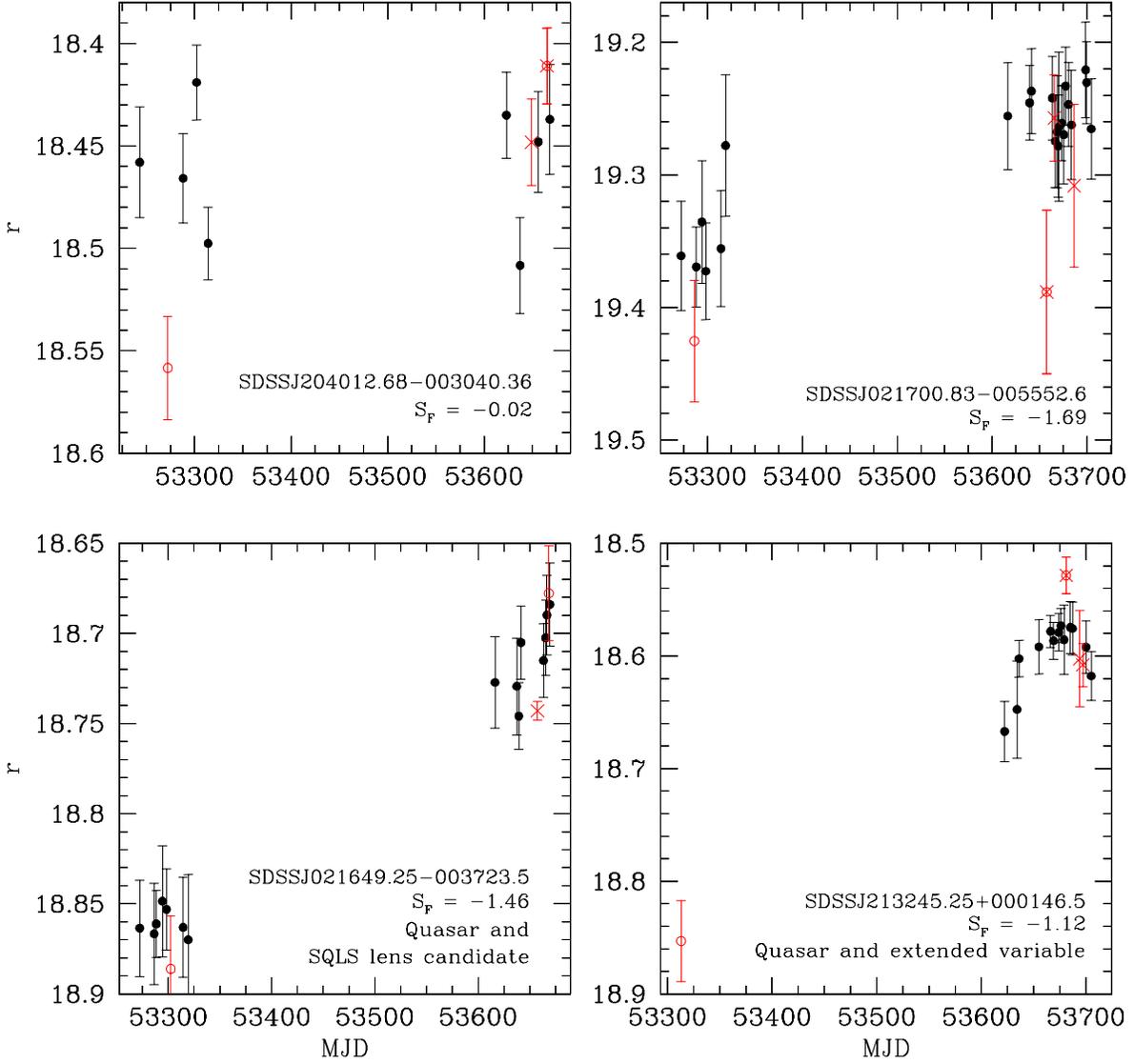}
\caption{Light curves of variable sources passing the $r$$-$$\sigma_r$ cut with different $S_F$ values.  Epochs that were cut are marked in red, with crosses meaning the epoch was on average more than 1 $\sigma$ away for all sources, and rings meaning the epoch was one of the two furthest from the mean for that particular source.  The sources on top are not known quasars, while the two on the bottom are SDSS spectroscopic quasars \citep{Schneider07}.}
\label{fig:LCs}
\end{figure}

In many ways, our experiment was limited by the data.  First, there are no known lensed quasars in Stripe 82, so we could evaluate no examples of success or set any limits on the role of false negatives.  Furthermore, when we carried out this experiment, the Supernova Survey had few public ($\la 25$) epochs available.  Since some epochs will be dropped because of bad seeing, or because image subtraction failed, there are some fields where there were too few enough epochs to extract light curve statistics.  Similarly, the released epochs did not cover a large enough time interval.  In some cases (e.g. SDSSJ213245.25+000146.5 in Figure~\ref{fig:LCs}), all the non-rejected epochs spanned only a few months.  The performance of our approach would improve markedly with survey data spanning several years and with many more epochs.  Additionally, the relatively poor resolution of the SDSS survey is not ideal for identifying quasar lenses.  This is also true for color and morphological techniques, such as in \citet{Inada07}, where candidate lenses with image separations of less than 1\farcs0 were rejected.  With better resolution and more epochs, false positives in the method would decrease.  If we examined every SDSS quasar in the supernova region, we would expect to find of order 1000 were variable with $\sim 50$ false positives if we assume no improvement in the false positive rate with the addition of more epochs.  In reality, we should have a marked reduction in the rate because we could focus on epochs with better seeing and add in the variability information from the ugiz bands as well. 

While these shortcomings will be partly solved by the full SDSS Supernova Survey data set, the real future for the method lies with ground-based surveys like LSST \citep{Tyson02} and Pan-STARRS \citep{Kaiser04} and proposed surveys from space like the SuperNova Acceleration Probe \citep[SNAP,][]{Aldering02}.  Pan-STARRS and LSST would repeatedly survey very large areas ($10^3$--$10^4$ rather than $10^2$ square degrees) with improved resolution ($0\farcs5$--$1\farcs$ rather than $>1\farcs0$) and depth ($\simeq 24$~mag rather than $\simeq 22$~mag).  \citet{Kochanek06} estimated that LSST can discover $\sim 10^3$ lensed quasars with V$<23$.  The space based surveys would cover far less area (15~square degree for SNAP) but with vastly improved resolution ($\simeq 0\farcs05$) and depth ($\simeq 28$~mag).  Since few lenses have separations this small (1\farcs5 is the expected median, e.g. \citet{Mitchell05}), any lens identified by variability is easily confirmed by its morphology.  More important for the space missions is that their great depth allows searches for other lensed, time variable sources such as the supernovae themselves.

\acknowledgments
    Funding for the Sloan Digital Sky Survey (SDSS) and SDSS-II has been provided by the Alfred P. Sloan Foundation, the Participating Institutions, the National Science Foundation, the U.S. Department of Energy, the National Aeronautics and Space Administration, the Japanese Monbukagakusho, and the Max Planck Society, and the Higher Education Funding Council for England. The SDSS Web site is http://www.sdss.org/.

    The SDSS is managed by the Astrophysical Research Consortium (ARC) for the Participating Institutions. The Participating Institutions are the American Museum of Natural History, Astrophysical Institute Potsdam, University of Basel, University of Cambridge, Case Western Reserve University, The University of Chicago, Drexel University, Fermilab, the Institute for Advanced Study, the Japan Participation Group, The Johns Hopkins University, the Joint Institute for Nuclear Astrophysics, the Kavli Institute for Particle Astrophysics and Cosmology, the Korean Scientist Group, the Chinese Academy of Sciences (LAMOST), Los Alamos National Laboratory, the Max-Planck-Institute for Astronomy (MPIA), the Max-Planck-Institute for Astrophysics (MPA), New Mexico State University, Ohio State University, University of Pittsburgh, University of Portsmouth, Princeton University, the United States Naval Observatory, and the University of Washington.

This research was supported by NSF grant AST-0708082.

N.~I. acknowledges support from the Special Postdoctoral Researcher Program of RIKEN. 

This work was supported in part by Department of Energy contract DE-AC02-76SF00515.

\begin{deluxetable}{lrrrr}
\tablecaption{SQLS LENS CANDIDATES IN THE SDSS SUPERNOVA SURVEY REGION}
\tablehead{\colhead{Lens Candidate} & \colhead{$\alpha_1$\tablenotemark{a}} & \colhead{$\delta_1$\tablenotemark{a}} & \colhead{$\alpha_2$\tablenotemark{a}} & \colhead{$\delta_2$\tablenotemark{a}}}
\startdata
SDSSJ0020$-$0011\tablenotemark{b} & 00\hmsh20\hmsm23\fs18 & $-$0\arcdeg11\arcmin10\farcs7 & \nodata & \nodata\\
SDSSJ0141+0031\tablenotemark{b} & 01\hmsh41\hmsm11\fs62 & 0\arcdeg31\arcmin44\farcs8 & 01\hmsh41\hmsm10\fs34 & 0\arcdeg31\arcmin07\farcs0\\
SDSSJ0212+0034\tablenotemark{c} & 02\hmsh12\hmsm49\fs60 & 0\arcdeg34\arcmin48\farcs7 & \nodata & \nodata\\
SDSSJ0213+0032\tablenotemark{b} & 02\hmsh13\hmsm23\fs24 & 0\arcdeg32\arcmin56\farcs9 & \nodata & \nodata\\
SDSSJ0216$-$0037\tablenotemark{c} & 02\hmsh16\hmsm49\fs26 & $-$0\arcdeg37\arcmin23\farcs6 & 02\hmsh16\hmsm49\fs16 & $-$0\arcdeg37\arcmin11\farcs5\\
SDSSJ0216$-$0102\tablenotemark{c} & 02\hmsh16\hmsm45\fs80 & $-$1\arcdeg02\arcmin04\farcs8 & \nodata & \nodata\\
SDSSJ0232+0106\tablenotemark{c} & 02\hmsh32\hmsm05\fs09 & 1\arcdeg06\arcmin40\farcs3 & 02\hmsh32\hmsm05\fs19 & 1\arcdeg06\arcmin34\farcs2\\
SDSSJ0248+0009\tablenotemark{b} & 02\hmsh48\hmsm20\fs78 & 0\arcdeg09\arcmin56\farcs5 & 02\hmsh48\hmsm21\fs41 & 0\arcdeg09\arcmin56\farcs9\\
SDSSJ0249+0025\tablenotemark{b} & 02\hmsh49\hmsm24\fs67 & 0\arcdeg25\arcmin36\farcs1 & \nodata & \nodata\\
SDSSJ0249+0039\tablenotemark{b} & 02\hmsh49\hmsm07\fs75 & 0\arcdeg39\arcmin16\farcs6 & \nodata & \nodata\\
SDSSJ0258$-$0010\tablenotemark{c} & 02\hmsh58\hmsm04\fs27 & $-$0\arcdeg11\arcmin00\farcs0 & 02\hmsh58\hmsm03\fs82 & $-$0\arcdeg11\arcmin18\farcs2\\
SDSSJ2038+0055\tablenotemark{c} & 20\hmsh38\hmsm45\fs36 & 0\arcdeg55\arcmin32\farcs1 & 20\hmsh38\hmsm46\fs09 & 0\arcdeg55\arcmin41\farcs4\\
SDSSJ2040$-$0030\tablenotemark{c} & 20\hmsh40\hmsm30\fs53 & $-$0\arcdeg30\arcmin15\farcs9 & 20\hmsh40\hmsm30\fs71 & $-$0\arcdeg30\arcmin10\farcs6\\
SDSSJ2052+0011\tablenotemark{b} & 20\hmsh52\hmsm12\fs82 & 0\arcdeg11\arcmin37\farcs5 & 20\hmsh52\hmsm13\fs85 & 0\arcdeg11\arcmin16\farcs4\\
SDSSJ2057+0006\tablenotemark{b} & 20\hmsh57\hmsm52\fs49 & 0\arcdeg06\arcmin35\farcs3 & \nodata & \nodata\\
SDSSJ2122$-$0026\tablenotemark{c} & 21\hmsh22\hmsm43\fs02 & $-$0\arcdeg26\arcmin53\farcs7 & \nodata & \nodata\\
SDSSJ2124$-$0047\tablenotemark{c} & 21\hmsh24\hmsm29\fs83 & $-$0\arcdeg47\arcmin27\farcs1 & 21\hmsh24\hmsm30\fs91 & $-$0\arcdeg47\arcmin25\farcs3\\
SDSSJ2129$-$0051\tablenotemark{c} & 21\hmsh29\hmsm56\fs45 & $-$0\arcdeg51\arcmin50\farcs5 & 21\hmsh29\hmsm56\fs57 & $-$0\arcdeg51\arcmin52\farcs5\\
SDSSJ2132+0000\tablenotemark{b} & 21\hmsh32\hmsm36\fs62 & 0\arcdeg00\arcmin17\farcs6 & 21\hmsh32\hmsm33\fs72 & 0\arcdeg00\arcmin09\farcs4\\
SDSSJ2134$-$0054\tablenotemark{c} & 21\hmsh34\hmsm14\fs02 & $-$0\arcdeg45\arcmin33\farcs1 & 21\hmsh34\hmsm14\fs18 & $-$0\arcdeg45\arcmin14\farcs7\\
SDSSJ2139$-$0114\tablenotemark{c} & 21\hmsh39\hmsm32\fs17 & $-$1\arcdeg14\arcmin05\farcs8 & \nodata & \nodata\\
SDSSJ2200$-$0107\tablenotemark{b} & 22\hmsh00\hmsm00\fs02 & $-$1\arcdeg07\arcmin48\farcs0 & \nodata & \nodata\\
SDSSJ2211$-$0009\tablenotemark{c} & 22\hmsh11\hmsm10\fs99 & $-$0\arcdeg09\arcmin53\farcs4 & \nodata & \nodata\\
SDSSJ2228$-$0059\tablenotemark{c} & 22\hmsh28\hmsm22\fs17 & $-$0\arcdeg59\arcmin43\farcs6 & 22\hmsh28\hmsm22\fs19 & $-$0\arcdeg59\arcmin49\farcs4\\
SDSSJ2337+0056\tablenotemark{c} & 23\hmsh37\hmsm13\fs67 & 0\arcdeg56\arcmin10\farcs9 & \nodata & \nodata\\
SDSSJ2351+0047\tablenotemark{b} & 23\hmsh51\hmsm48\fs36 & 0\arcdeg47\arcmin51\farcs6 & \nodata & \nodata\\
\enddata
\tablenotetext{a}{Component 1 is always a spectroscopically identified quasar given in \citet{Schneider05}.  If the candidate images are less than 2\farcs5 apart, the components are blended, and no position is given for Component 2.  If they are more than 2\farcs5 apart, Component 2 is selected by color criteria and the position is given.}
\tablenotetext{b}{Positions from initial SQLS candidate list, but these candidates were outside the final selection criteria for \citet{Inada07}.}
\tablenotetext{c}{Positions listed in \citet{Inada07}.}
\label{table:InadaLenses}
\end{deluxetable}

\begin{deluxetable}{lrrrr}
\tablecaption{NUMBER OF SOURCES SATISFYING EACH CUT}
\tablehead{\colhead{Criteria} & \colhead{ISIS sources} & \multicolumn{2}{c}{Candidate lensed quasars\tablenotemark{a}} & \colhead{SDSS QSOs\tablenotemark{b}} \\ & & Systems & Images & }
\startdata
In the fields & \nodata\3phn & 26\2phn & 39\2phn & 60\4phn\\
Detected by ISIS & 20768\3phn & 25\2phn & 36\2phn & 49\4phn\\
Have light curve statistics & 15065\3phn & 21\2phn & 29\2phn & 44\4phn\\
Variable in $r$$-$$\sigma_r$\tablenotemark{c} & 370\3phn & 14\2phn & 14\2phn & 25\4phn\\
Pass F-test\tablenotemark{d} & 21\3phn & 8\2phn & 8\2phn & 15\4phn\\
Measured on ``absdiff'' images & 18\3phn & 7\2phn & 7\2phn & 14\4phn\\
Variable on ``absdiff'' images\tablenotemark{e} & 18\3phn & 7\2phn & 7\2phn & 14\4phn\\
Elongated or large\tablenotemark{f} & 1\3phn & 0\2phn & 0\2phn & 1\4phn\\
\cutinhead{With F-test and Ignoring $r$$-$$\sigma_r$ Variability Cut}
Pass F-test, no $r$$-$$\sigma_r$ cut\tablenotemark{d} & 208\3phn & 8\2phn & 8\2phn & 16\4phn\\
Measured on ``absdiff'' images & 85\3phn & 7\2phn & 7\2phn & 15\4phn\\
Variable on ``absdiff'' images\tablenotemark{e} & 54\3phn & 7\2phn & 7\2phn & 15\4phn\\
Elongated or large\tablenotemark{f} & 15\3phn & 0\2phn & 0\2phn & 1\4phn\\
\cutinhead{With $r$$-$$\sigma_r$ Variability Cut and Ignoring F-test}
Variable $r$$-$$\sigma_r$, no $S_F$ cut\tablenotemark{c} & 370\3phn & 14\2phn & 14\2phn & 25\4phn\\
Measured on ``absdiff'' images & 311\3phn & 13\2phn & 13\2phn & 24\4phn\\
Variable on ``absdiff'' images\tablenotemark{e} & 158\3phn & 12\2phn & 12\2phn & 22\4phn\\
Elongated or large\tablenotemark{f} & 64\3phn & 1\2phn & 1\2phn & 3\4phn\\
\enddata
\tablenotetext{a}{As in the initial SQLS candidate list \citep{Inada07}.}
\tablenotetext{b}{As given by \citet{Schneider07}.}
\tablenotetext{c}{Determined by $r$-$\sigma_r$ cut listed in \S 2.1, for sources with statistics}
\tablenotetext{d}{Determined by $S_F \le -0.7$}
\tablenotetext{e}{Determined by $\mu < \mu_{back} + 0.5$}
\tablenotetext{f}{Determined by $\epsilon > 1.75$ or $R_e > 2.5$~pixel}
\label{table:SourceNumbers}
\end{deluxetable}

\begin{deluxetable}{llrrrrr}
\tablecaption{CANDIDATE EXTENDED VARIABLE SOURCES}
\tablehead{\colhead{Object} & \colhead{Field} & \colhead{m} & \colhead{$\sigma_m$} & \colhead{$S_F$} & \colhead{$\epsilon$} & \colhead{$R_e$\tablenotemark{a}}}
\startdata
\cutinhead{Pass $r$$-$$\sigma_r$ Variability Cut}
SDSSJ213245.25+000146.5 & SDSSJ2132+0000 & 18.60 & 0.028 & $-$1.117 & 1.153 & 2.788 \\
\cutinhead{Fail $r$$-$$\sigma_r$ Variability Cut}
SDSSJ002016.45$-$000329.6 & SDSSJ0020$-$0011 & 19.09 & 0.020 & $-$0.926 & 1.222 & 3.054 \\
SDSSJ002035.77$-$000536.3 & SDSSJ0020$-$0011 & 20.47 & 0.031 & $-$0.765 & 1.999 & 0.644 \\
SDSSJ021247.22+003433.5 & SDSSJ0212+0034 & 17.56 & 0.009 & $-$0.802 & 1.765 & 12.181 \\
SDSSJ021315.87+003039.1 & SDSSJ0213+0032 & 19.91 & 0.020 & $-$0.827 & 1.425 & 4.569 \\
SDSSJ021328.83+003030.8 & SDSSJ0213+0032 & 18.57 & 0.013 & $-$0.770 & 1.679 & 3.995 \\
SDSSJ021639.19$-$002951.5 & SDSSJ0216$-$0037 & 19.81 & 0.024 & $-$0.990 & 1.274 & 2.547 \\
SDSSJ021655.74$-$005536.0 & SDSSJ0216$-$0102 & 20.38 & 0.049 & $-$0.944 & 1.438 & 4.315 \\
SDSSJ024805.87+000750.3 & SDSSJ0248+0009 & 18.64 & 0.016 & $-$0.775 & 1.337 & 3.829 \\
SDSSJ024904.31+004414.9 & SDSSJ0249+0039 & 19.44 & 0.020 & $-$1.130 & 1.312 & 2.702 \\
SDSSJ213228.60+000859.0 & SDSSJ2132+0000 & 19.22 & 0.023 & $-$0.718 & 2.126 & 2.104 \\
SDSSJ221056.01$-$000432.7 & SDSSJ2211$-$0009 & 19.71 & 0.033 & $-$0.758 & 1.341 & 3.593 \\
SDSSJ221102.93$-$000518.3 & SDSSJ2211$-$0009 & 19.04 & 0.006 & $-$0.712 & 2.661 & 5.878 \\
SDSSJ221122.22$-$000322.6 & SDSSJ2211$-$0009 & 15.61 & 0.007 & $-$0.746 & 1.019 & 2.557 \\
SDSSJ221119.29$-$000115.5 & SDSSJ2211$-$0009 & 18.89 & 0.012 & $-$0.825 & 1.770 & 1.894 \\
\enddata
\tablenotetext{a}{Given in units of pixels.  The image scale is 0\farcs396 pixel$^{-1}$.}
\label{table:ExtVarCands}
\end{deluxetable}

\end{document}